\documentclass[pra,superscriptaddress,twocolumn]{revtex4}
\usepackage{amsfonts}
\usepackage{amsmath}
\usepackage{amsthm}
\usepackage{amscd}
\usepackage{amssymb}
\usepackage{subfigure}
\usepackage{amsxtra}
\usepackage{bm}           
\usepackage{bbm}
\usepackage{graphicx}
\usepackage{epstopdf}
\usepackage{color}

\def\bi#1\ei {\begin{itemize}#1\end{itemize}}
\def\bn#1\en {\begin{enumerate}#1\end{enumerate}}
\def\bea#1\eea {\begin{align}#1\end{align}}
\def\bean#1\eean {\begin{align*}#1\end{align*}}
\def\ben#1\een {\begin{equation*}#1\end{equation*}}
\def\be#1\ee {\begin{equation}#1\end{equation}}
\def\bes#1\ees {\begin{equation}\begin{split}#1\end{split}\end{equation}}
\def\bear#1\eear {\begin{eqnarray}#1\end{eqnarray}}
\def\bear#1\eear {\begin{eqnarray*}#1\end{eqnarray*}}

\newcommand{\beq}{\begin{equation}}
\newcommand{\eeq}{\end{equation}}

\newcommand{\ket}[1]{\ensuremath{\left|#1\right\rangle}}
\newcommand{\braket}[2]{\ensuremath{\langle #1|#2\rangle}}
\newcommand{\ketbra}[2]{\ensuremath{| #1 \rangle \langle #2 |}}

\newcommand{\Hil}{\mathcal{H}}

\renewcommand{\qed}{\ensuremath{\hfill \blacksquare}}
\newcommand{\order}[1]{\mathcal{O}(#1)}
\newcommand{\alpsi}{\ensuremath{\ket{\alpha,\psi}}}

\newtheorem{thm}{Theorem}

\newtheorem{obs}[thm]{Observation}

\begin{document}

\title{Quantum Communication with Coherent States and Linear Optics}

\author{Juan Miguel Arrazola}
\author{Norbert L\"utkenhaus}

\affiliation{Institute for Quantum Computing and Department of Physics and Astronomy,
University of Waterloo, 200 University Avenue West,
N2L 3G1 Waterloo, Ontario, Canada}

\begin{abstract}
We introduce a general mapping for encoding quantum communication protocols involving pure states of multiple qubits, unitary transformations, and projective measurements into another set of protocols that employ coherent states of light in a superposition of optical modes, linear optics transformations and measurements with single-photon threshold detectors. This provides a general framework for transforming protocols in quantum communication into a form in which they can be implemented with current technology. We explore the similarity between properties of the original qubit protocols and the coherent-state protocols obtained from the mapping and make use of the mapping to construct new protocols in the context of quantum communication complexity and quantum digital signatures. Our results have the potential of bringing a wide class of quantum communication protocols closer to their experimental demonstration.
\end{abstract}

\date{\today}

\maketitle
What information-processing tasks are unachievable in a classical world but become possible when exploiting the intrinsic quantum mechanical properties of physical systems? This question has been a driving force of numerous research endeavours over the last few decades and remarkable progress has been made in our understanding of the advantages that quantum mechanics can provide, as well is in developing the experimental platforms that will allow them to be realized in practice \cite{ladd2010quantum,RevModPhys.82.665,giovannetti2006quantum,scarani2009security}. An example is the field of quantum communication \cite{gisin2007quantum}, where quantum systems can be used, for instance, to distribute secret keys \cite{bennett92b,bennett84a} or reduce the amount of communication required for joint computations \cite{QuantumFingerprinting,DJexp,RazProblem,HM-Bar-Yossef}.

In terms of experimental implementations, only quantum key distribution (QKD) has been routinely demonstrated and deployed over increasingly complex networks and large distances \cite{stucki2009high,sasaki2011field}. This is possible largely due to the fact that, fundamentally, QKD can be carried out with sequences of independent signals and measurements \cite{scarani2009security}. Other tasks typically require sophisticated quantum states to be prepared and transmitted as well as performing complex operations on them. Overall, there is a large set of quantum communication protocols whose potential advantages currently escape the grasp of available technology, with only a few proof-of-principle implementations having been reported \cite{ZhangQCC,TrojekQCC,HornFPs}.

Confronted with these challenges we face two alternatives: We can either strive to improve current technology or we can flip the issue around and ask: Can protocols in quantum communication be adapted to a form that makes them ready to be deployed with available techniques? To adopt the latter strategy is to push for a theoretical reformulation that converts previously intractable protocols into a form that, while conserving their relevant features, eliminates the obstacles affecting their implementation. This is precisely the road that has already been successfully followed for QKD.

In this work, we describe an abstract mapping that converts quantum communication protocols that use pure states of multiple qubits, unitary operations, and projective measurements into another class of protocols that use only a sequence of coherent states, linear optics operations, and measurements with single-photon threshold detectors. The new class of protocols requires a number of optical modes equal to the dimension of the original states, but the total number of photons can be chosen independently from the dimension and is typically very small. The protocols obtained from the mapping share important properties with the original ones, meaning that they can also fulfil the goal that the original protocols where intended to achieve. Overall, the mapping is suitable for its application to protocols that originally require a moderate number of qubits, but are still hard to implement with usual methods.

In the remainder of this paper, we describe the mapping in detail and discuss the various properties of the coherent-state protocols. We proceed by examining how the mapping can be applied to construct protocols in quantum communication complexity and describe protocols for the hidden matching problem and for quantum digital signatures, both of which can be realized with technology that is within current reach.

\section{Coherent-state mapping}\label{CSProtocols}

We consider a wide class of quantum communication protocols that require only three basic operations: the preparation of pure states of a fixed dimension, unitary transformations on these states, and projective measurements on a canonical basis. The simplest form of a protocol in this class is one in which Alice prepares a state $\ket{\psi}$ and sends it to Bob, who then applies a unitary transformation $U_B$ to that state, followed by a projective measurement on the canonical basis. More complex protocols can be constructed by increasing the number of these basic operations as well as the number of parties. Even though these protocols generally involve states of some arbitrary dimension $d$, it is common to think of them as corresponding to a system of $\order{\log_2 d}$ qubits. Hence, we refer to them as \textit{qubit protocols}.

An \textit{exact} implementation of such protocols can be achieved without the use of actual physical qubits by instead considering a single photon in a superposition of optical modes. Any pure state $\ket{\psi}=\sum_{k=1}^d \lambda_{k} \ket{k}$, with $\sum_{k=1}^d |\lambda_{k}|^2=1$, can be equally thought of as the state of a single photon in a superposition of $d$ modes
\beq\label{single-photon}
a_{\psi}^{\dagger}\ket{0}=\sum_{k=1}^d \lambda_{k} \ket{1}_k,
\eeq
where $a^{\dagger}_{\psi}=\sum_{k=1}^d \lambda_{k} b^{\dagger}_k$ for a collection of creation operators $\{b_1,b_2,\ldots,b_d\}$ corresponding to $d$ optical modes, and where $\ket{1}_k$ is the state of a single photon in the $k-$th mode. 

In this picture, unitary operations correspond exactly to linear optics transformations \cite{reck94a}, and measurements in the canonical basis are equivalent to a photon counting measurement in each of the modes. 

From a practical perspective, the issue with implementing qubit protocols in terms of a single photon in a superposition of modes is that the experimental preparation of these states presents daunting challenges. Instead, we are interested in an adaptation of this formulation of qubit protocols into another that is more readily implementable in practice. As discussed in Ref. \cite{massar2005quantum}, as an alternative to a single photon we can consider a coherent state in a superposition of modes. In that case, instead of the state of Eq. \eqref{single-photon}, we have
\beq\label{alpsi}
D_{a_{\psi}}(\alpha)\ket{0}=\bigotimes_{k=1}^d\ket{\alpha\,\lambda_{k}}_k,
\eeq
where $D_{a_{\psi}}(\alpha)=\exp(\alpha a_{\psi}^{\dagger}-\alpha^* a_{\psi} )$ is the displacement operator. Note that this state is simply a sequence of coherent states over $d$ optical modes.

With this idea in mind, we now outline a method for converting qubit protocols into another class of protocols that, although seemingly disparate, actually retain the essential properties of the original ones. We call these \textit{coherent-state protocols} since they can be implemented by using only coherent states of light and linear optics operations. The recipe for constructing coherent-state protocols is specified by the following rules:\\\\\\

\textbf{Coherent-state mapping}

\begin{enumerate}
\item{The original Hilbert space $\Hil$ of dimension $d$ with canonical basis $\{\ket{1},\ket{2},\ldots,\ket{d}\}$ is mapped to a set of $d$ orthogonal optical modes with corresponding annihilation operators $\{b_1,b_2,\ldots,b_d\}$:
\beq
\ket{k}\longrightarrow b_k.
\eeq}
\item{A state $\ket{\psi}=\sum_{k=1}^d \lambda_{k} \ket{k}$ is mapped to a coherent state with parameter $\alpha$ in the mode $a_{\psi}=\sum_{k=1}^d \lambda_{k} b_k$: 
\begin{align}
\ket{\psi}\longrightarrow \ket{\alpha,\psi_x}&:=\bigotimes_{k=1}^d\ket{\alpha\,\lambda_{k}}_k,\nonumber
\end{align}
where $\ket{\alpha\,\lambda_{k}}_k$ is a coherent state with parameter $\alpha\,\lambda_{k}$ in the $k$-th mode. The value of the amplitude $\alpha$ can be chosen freely as a parameter of the mapping, independently of the dimension $d$, but remains fixed. } 
\item{A unitary operation $U$ acting on a state in $\Hil$ is mapped into linear optics transformation corresponding to the same unitary operator $U$ acting on the modes $\{b_1,b_2,\ldots,b_d\}$. Thus, the transformation of a state is linked to a transformation of the modes as:
\beq
\ket{\psi'}=U\ket{\psi}\longrightarrow \left( \begin{array}{c}
b_1 \\
b_2 \\
\vdots\\
b_d \end{array} \right) = U \left( \begin{array}{c}
b_1' \\
b_2'\\
\vdots\\
b_d'\end{array} \right).
\eeq}
\item{A projective measurement in the canonical basis $\{\ket{1},\ket{2},\ldots,\ket{d}\}$ is mapped into a two-outcome measurement in each of the modes with single-photon threshold detectors:
\beq
\{\ketbra{1}{1},\ketbra{2}{2},\ldots,\ketbra{d}{d}\}\longrightarrow \left\{\bigotimes_{k=1}^d F_{\text{c}}^k\right\},
\eeq
where $c=\text{``click", ``no-click"}$, $F_{\text{no-click}}^k=\ketbra{0}{0}$ is a projection onto the vacuum, $F_{\text{no-click}}^k=\sum_{n=1}^{\infty}|n\rangle_k\langle n|_k$, and $\ket{n}_k$ is a state with $n$ photons in the $k-$th mode. As such, an outcome in a coherent-state protocol corresponds to a pattern of clicks across the modes.}
\end{enumerate}

\begin{figure}
\includegraphics[width=\columnwidth]{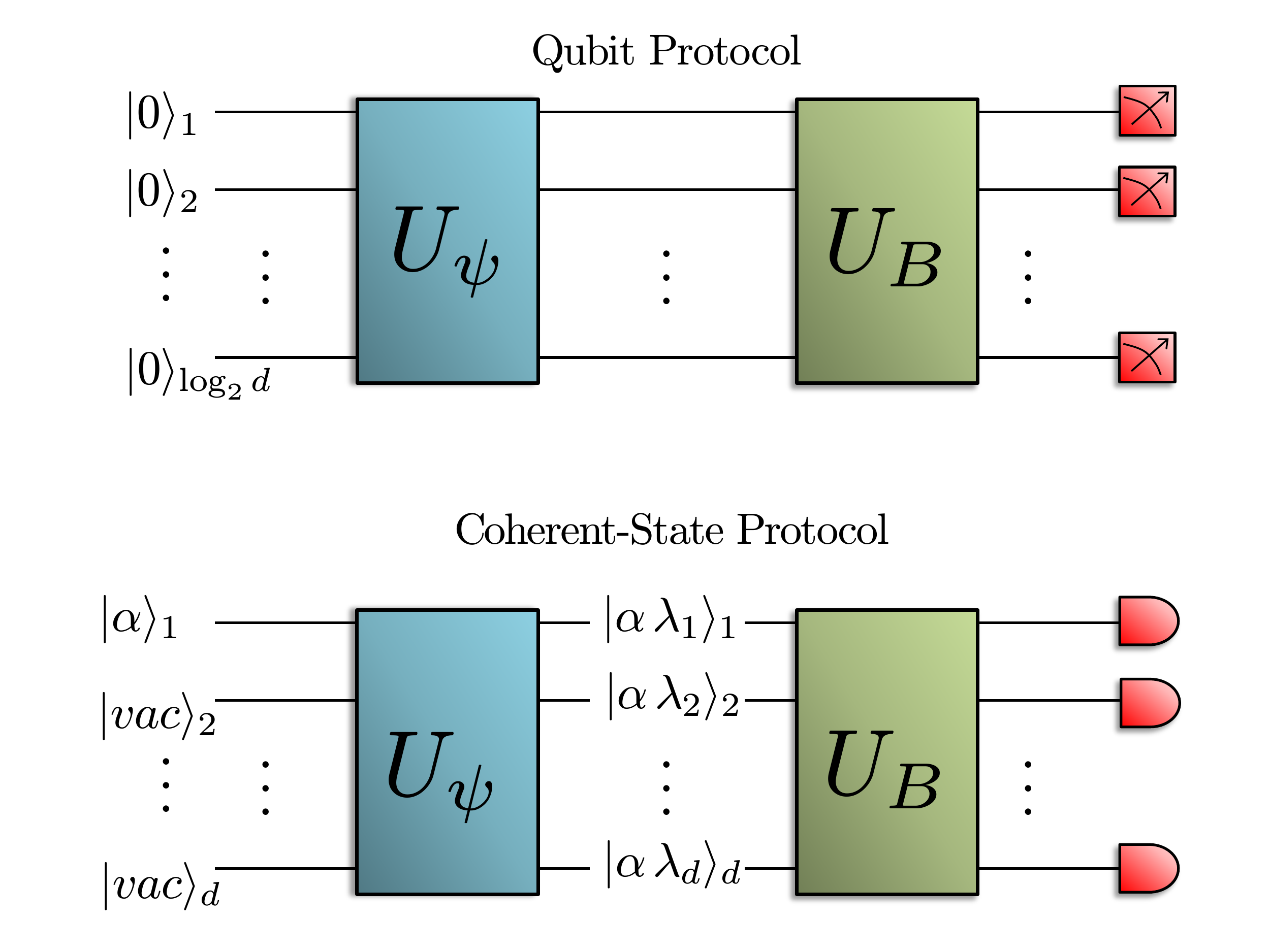}
\caption{(Color online) In a simple qubit protocol, Alice prepares a state $\ket{\psi}=\sum_{k=1}^{d}\lambda_k\ket{k}$ of $ \log_2 d$ qubits by applying a unitary transformation $U_{\psi}$ on an inital state $\ket{\bar{0}}:=\ket{0}^{\otimes \log_2 d}$. She sends the state to Bob, who applies a unitary transformation $U_B$ and measures the resulting state in the computational basis. In the equivalent coherent-state protocol, the initial state corresponds to a coherent state in a single mode and the vacuum on the others. The state $\ket{\alpha,\psi_x}=\bigotimes_{k=1}^d\ket{\alpha\,\lambda_{k}}_k$ is prepared by applying the transformation $U_{\psi}$ to the optical modes. This state is sent to Bob, who applies the transformation $U_B$ and consequently measures each mode for the presence of photons with threshold single-photon detectors.}\label{Mapping}
\end{figure}

Since any qubit protocol can be constructed from the basic operations of state preparation, unitary transformations, and projective measurements, the above instructions are sufficient to construct the coherent-state version of any qubit protocol. However, as there are $2^d$ possible outcomes compared to the $d$ possible outcomes of the qubit protocol, the interpretation of the outcomes in the coherent-state protocol is not immediately provided by the mapping. Nevertheless, as will be discussed later, the statistics closely resemble those of the original protocol and they can be thought of as arising from several independent runs of the original qubit protocol. As an illustration, a simple qubit protocol and its coherent-state counterpart are depicted in Fig. \ref{Mapping}.

An immediate appealing property of coherent-state protocols is that their implementation faces much lesser obstacles than their qubit counterparts. Indeed, the fundamental challenge of a quantum-optical implementation of qubit protocols lies in the difficulty of generating entangled states of many qubits and performing global unitary transformations on them. On the other hand, coherent-state protocols face significantly less daunting obstacles. The experimental generation of coherent states is a commonplace task and the construction of linear-optical circuits can, in principle, be realized with simple devices such as beam splitters and phase-shifters \cite{reck94a}, though experimental challenges may remain depending on the required unitary operation. Moreover, the platforms for linear optics experiments continue to improve at a fast rate, most notably with the development of integrated optics \cite{tanzilli2012genesis}.

As we have mentioned already, an advantage of coherent-state protocols is that they employ a coherent state in a superposition of modes, which is equivalent to a tensor product of individual coherent states across the various modes. However, qubit protocols usually require high amounts of entanglement. This seems to indicate that the `quantumness' of the original qubit protocol has been lost through the mapping. Nevertheless, it is important to realize that this is not the case, as coherent-state protocols showcase a truly quantum property: non-orthogonality. Given two states $\alpsi=\bigotimes_{k=1}^d\ket{\alpha\,\lambda_{k}}_k$ and $\ket{\alpha,\varphi}=\bigotimes_{k=1}^d\ket{\alpha\,\nu_{k}}_k$, with $d\gg \alpha$, the individual coherent states in each mode will typically be highly non-orthogonal, i.e. $\langle\alpha\,\nu_{k}|\alpha\,\lambda_k\rangle\approx 1$. In fact, it can be useful to intuitively think of the coherent-state mapping as an exchange between entanglement and non-orthogonality, since an implementation of qubit protocols with actual physical qubits usually requires entanglement amongst the qubits.

In coherent-state protocols, the average photon number, $|\alpha|^2$, is a parameter that can be chosen independently of the dimension of the states of the original qubit protocol. This is to be put in contrast with any quantum protocol that encodes a qubit in the degrees of freedom of a photon, which inevitably requires a number of photons that scales with the dimension of the states. Hence, coherent-state protocols offer an intrinsic saving in the number of photons required for their implementation. The drawback, of course, is that the required number of optical modes is equal to the dimension of the states in the original protocol.

Fortunately, current laser sources can operate with clock rates of at least 1GHz \cite{dixon2008gigahertz}, permitting the generation of a very large number of modes per second. This makes it possible in practice to apply the mapping to protocols involving states of a moderate number of qubits, whose dimension is comparable to the number of modes that can be manipulated with available devices. As we will discuss in sections \ref{QCC} and \ref{QDSsection}, there are many qubit protocols to which we can apply the mapping that require only a modest number of qubits but still currently escape the grasp of direct implementations. From a theoretical perspective, the relationship between these two types of protocols may provide an insight into the trade-offs between different resources in quantum communication, as well as into the interplay between entanglement and non-orthogonality in quantum mechanics.

Now that we have specified how to construct coherent-state protocols, our goal is to understand their properties. We pay special attention to their resemblance to qubit protocols, but also concentrate on understanding the features that may provide an advantage over their qubit counterparts or find applications in quantum communication.

\textit{Transmitted information.-} We are often interested in quantifying the amount of transmitted information, i.e. amount of communication, that takes place in a quantum protocol. Informally, this is done by counting the number of qubits that are employed. But what happens if a protocol uses physical systems that are manifestly \textit{not} qubits? In that case, we quantify the transmitted information in terms of the smallest number of qubits that would be required, in principle, to replicate the performance of the protocol. More precisely, if a quantum protocol uses states in a Hilbert space of dimension $d$, this space can be associated with a system of $\order{\log_2 d}$ qubits. Therefore, the amount of communication $C$ in a quantum protocol is generally given by
\beq
C=\log_2[\dim(\Hil)],
\eeq
where $\Hil$ is the smallest Hilbert space containing all states of the protocol, which can be a significantly smaller than the entire Hilbert space associated with the physical systems. Moreover, Holevo's theorem \cite{holevo1973bounds} guarantees that no more than $\order{\log_2 d}$ classical bits of information could be transmitted, on average, by a quantum protocol that uses states in a Hilbert space of dimension $d$.

By quantifying the amount of communication carefully, we gain a better understanding of the different physical resources that are required to transmit a certain amount of information. For example, the fact that the same amount of information can be transmitted by a single photon in $n$ optical modes, at most $n$ photons in a single mode or $\log_2 n$ qubits, is understood because the smallest Hilbert space containing all possible states in each of the three cases has the same dimension.

Quantifying the amount of transmitted information in qubit protocols is straightforward. For coherent-state protocols obtained from the mapping, even though the \textit{actual} Hilbert space associated with all possible signal states is large (distinct coherent states are linearly independent), they effectively occupy a small Hilbert space, as is expressed in the following theorem: 
\begin{thm}\label{dimensionthm}
\cite{arrazolaqfp} For any state $\ket{\psi}$ in a Hilbert space of dimension $d$ and for any $\epsilon>0$, there exists a Hilbert space $\Hil_{\alpha}$ of dimension $d_{\alpha}$ such that
\begin{align*}
&\langle\alpha,\psi|P_{\Hil_{\alpha}}|\alpha,\psi\rangle\geq 1-\epsilon,\\
&\log_2 d_{\alpha}=\order{\log_2 d},
\end{align*} 
and where $P_{\Hil_{\alpha}}$ is the projector unto $\Hil_{\alpha}$.
\end{thm}
\textit{Proof:} For a given $\Delta>0$, we choose $\Hil_{\alpha}$ to be the subspace spanned by the set of Fock states $\{\ket{n_1}\otimes\ket{n_2}\otimes\ldots\otimes \ket{n_d}\}$ over $d$ modes whose total photon number $n=\sum_{k=1}^dn_k$ satisfies $|n-|\alpha|^2|\leq\Delta$. In other words, this is the space of states whose total photon number is close to $|\alpha|^2$.

The dimension of the Hilbert space spanned by states of $n$ photons is equal to the the number of distinct ways in which $n$ photons can be distributed into the $d$ different modes. Since the photons are indistinguishable, this quantity is given by the binomial factor $\binom{n+d-1}{d-1}$ \cite{2002first}. In the case of $\Hil_{\alpha}$, there are $2\Delta$ different possible values of $n$, the largest being $n=|\alpha|^2 +\Delta$. Thus, the dimension $d_{\alpha}$ of this subspace satisfies
\beq
d_{\alpha}\leq 2\Delta\binom{|\alpha|^2 +\Delta+d-1}{d-1},
\eeq
which gives
\begin{align}
&\log_2 d_{\alpha}\leq \log_2\left[2\Delta\binom{|\alpha|^2 +\Delta+d-1}{d-1}\right]\nonumber\\
\leq&(|\alpha|^2+\Delta) \log_2\left[(|\alpha|^2 +\Delta+d-1)\right]+\log_2(2\Delta),
\end{align}
which is $\order{\log_2 d}$ for any fixed $\alpha$ and $\Delta$. 

Now notice that the number $\langle\alpha,\psi|P_{\Hil_{\alpha}}|\alpha,\psi\rangle$ is equal to the probability of performing a photon number measurement on $\alpsi$ and obtaining a value $n$ satisfying $|n-|\alpha|^2|\leq\Delta$. Since any coherent state $\alpsi$ has a Poissonian photon number distribution with mean $|\alpha|^2$, independently of $\ket{\psi}$, we can use the properties of this distribution to calculate the probability that the measured number of photons lies within the desired range. This probability satisfies \cite{franceschetti2007closing}
\begin{align}
&P(|n-|\alpha|^2|\geq\Delta)\leq 2e^{-|\alpha|^2}\left(\frac{e|\alpha|^2}{|\alpha|^2+\Delta}\right)^{|\alpha|^2+\Delta}\label{BoundPoisson}
\end{align}
which can be made equal to any $\epsilon>0$ by choosing $\Delta$ accordingly while keeping $\alpha$ fixed. \qed

Therefore, the fact that the mean photon number $|\alpha|^2$ is fixed in coherent-state protocols leads to the states involved effectively occupying a Hilbert space of dimension that is comparable to that of the original one. This implies that the asymptotic behaviour of the amount of transmitted information is the same for both classes of protocols. Moreover, the effectively unused sections of the entire Hilbert space can still be used, in principle, for other purposes such as the transmission of additional classical or quantum information through multiplexing schemes. A method for achieving this in practice is a line for future research.

It is important to note that this correspondence in the transmitted information is not exactly mirrored in terms of the expenditure of physical resources. A coherent-state protocol obtained from the mapping employs $d$ modes but a number of photons that is tunable and independent of this dimension. This is to be put in contrast with any quantum protocol that encodes a qubit in the degrees of freedom of a photon, which employs $\order{\log_2 d}$ optical modes and $\order{\log_2 d}$ photons.

\textit{Outcome probabilities.-} In qubit protocols, the probability of obtaining an outcome $k$ upon a measurement of a state $\ket{\psi}=\sum_{k=1}^d\lambda_k\ket{k}$ is given by
\beq
p_k=|\langle k|\psi\rangle|^2=|\lambda_k|^2,
\eeq 
with $\sum_{k=1}^d p_k=1$. For coherent-state protocols, the situation is different since we are performing independent measurements on each of the modes. In this case, the individual detector clicks are not mutually exclusive: We can have many clicks across the various modes, or even no clicks at all. 

Nevertheless, for the state $\alpsi$, the probability distribution of the number of photons in each mode is equivalent to the one obtained from many repetitions of a measurement on the single-photon state $\ket{\psi}=\alpha_{\psi}^{\dagger}\ket{0}$ of Eq. \eqref{single-photon}, where the number of repetitions is drawn from a Poisson distribution with mean $\mu=|\alpha|^2$.

To see this, first note that the state $\alpsi$ can be written as
\begin{align}
\alpsi&=D_{a_{\psi}}(\alpha)\ket{0}\nonumber\\
&=e^{-|\alpha|^2}\sum_{n=0}^{\infty}\frac{\alpha^n}{\sqrt{n!}}\ket{n}_{a_{\psi}}.
\end{align}
The state of $n$ photons in mode $\alpha_{\psi}$ is itself given by
\beq\label{alpsi}
\ket{n}_{a_{\psi}}=\frac{1}{\sqrt{n!}}(\alpha_{\psi}^{\dagger})^n\ket{0}=\frac{1}{\sqrt{n!}}\left(\sum_{k=1}^d\lambda_kb_k^{\dagger}\right)^n\ket{0}.
\eeq
For this state, the probability of obtaining $n_1,n_2,\ldots, n_d$ photons in each of the modes $b_1,b_2,\ldots,b_d$, with $\sum_k n_k=n$, is given by
\beq
\Pr(n_1,\ldots, n_d)=\frac{n!}{n_1!\ldots n_d!}|\lambda_1|^{2n_1}\ldots|\lambda_d|^{2n_d},
\eeq
which, from the multinomial theorem, is exactly equal to that obtained from $n$ measurements of the single-photon state $\ket{\psi}=\alpha^{\dagger}_{\psi}\ket{0}$. Since the number of photons $n$ in the state $\alpsi$ are Poissonian distributed with mean $\mu=|\alpha|^2$, this proves the claim. 

Whenever possible, we will not use photon-number resolving detectors in protocols obtained from coherent-state mapping, but threshold detectors that give clicks or no clicks. Note that while the statistics of photon counts is directly derived from the Poissonian distribution of repetitions of the single-photon protocol, this does not hold for the statistics of clicks of the threshold detectors.

However, for most states, the coefficients $\lambda_k$ will typically be very small, so that the mean number of photons $|\alpha\, \lambda_k|^2$ will also be small provided $\alpha$ is not too large. Then it is unlikely that more than one photon will be present in each mode, and the number-resolving properties of the detectors are not necessary. 

For example, with threshold detectors, the probability of obtaining a click on the $k$-th mode after a measurement of a state $\alpsi=\bigotimes_{k=1}^d\ket{\alpha\,\lambda_{k}}_k$ is given by
\beq
p_{\alpha,k}=1-\exp(-|\alpha\,\lambda_k|^2),
\eeq
which for $|\alpha\,\lambda_k|\ll 1$ gives
\beq
p_{\alpha,k}\approx |\alpha\,\lambda_k|^2.
\eeq
If we choose $|\alpha|^2=1$, we recover a behaviour very similar to that of the qubit protocol: Only one click is expected to occur and it does so with a probability that is essentially identical to that of the original protocol.

In any case, the multiple-photon property of a coherent-state protocol constitutes a potential advantage over its qubit counterpart. The expected number of clicks can be controlled by modifying $\alpha$ appropriately, and a larger number of clicks will give rise to more information gained per measurement.

\textit{State overlap.-} All of the physically-relevant information of a set of quantum states $\{\ket{\psi_1},\ket{\psi_2},\ldots,\ket{\psi_N}\}$ is contained in its Gram matrix, which is defined as
\beq
G_{i,j}=\braket{\psi_i}{\psi_j}.
\eeq
Thus, for quantum communication protocols defined over a set of possible signal states, it is natural to ask how the overlap between states behaves under a coherent-state mapping. The answer is provided by the following observation.
\begin{obs}
Let $\ket{\psi}=\sum_{k} \lambda_k\ket{k}$ and $\ket{\varphi}=\sum_k \nu_k\ket{k}$ be two arbitrary states with overlap $\braket{\psi}{\varphi}=\delta$. Then the overlap of their coherent-state versions satisfies
\beq
\delta_{\alpha}:=\braket{\psi,\alpha}{\varphi,\alpha}=\exp\left[|\alpha|^2(\braket{\psi}{\varphi}-1)\right].
\eeq
\end{obs}
\textit{Proof:} The overlap of the coherent-state versions is given by
\begin{align*}
&\braket{\psi,\alpha}{\varphi,\alpha}=\prod_{k}\braket{\alpha\,\lambda_k}{\alpha\,\nu_k}\\
&=\prod_{k}\exp\left[-\tfrac{|\alpha|^2}{2}(|\lambda_k|^2+|\nu_k|^2-2\lambda_k^*\nu_k)\right]\\
&=\exp\left[-\tfrac{|\alpha|^2}{2}\sum_{k}(|\lambda_k|^2+|\nu_k|^2-2\lambda_k^*\nu_k)\right]\\
&=\exp\left[|\alpha|^2(\braket{\psi}{\varphi}-1)\right],
\end{align*}
where we have used the relations $\sum_k |\lambda_k|^2=\sum_k |\nu_k|^2=1$ and $\braket{\psi}{\varphi}=\sum_k \lambda_k^*\nu_k$. \qed 

\begin{figure}
\includegraphics[width=0.9\columnwidth]{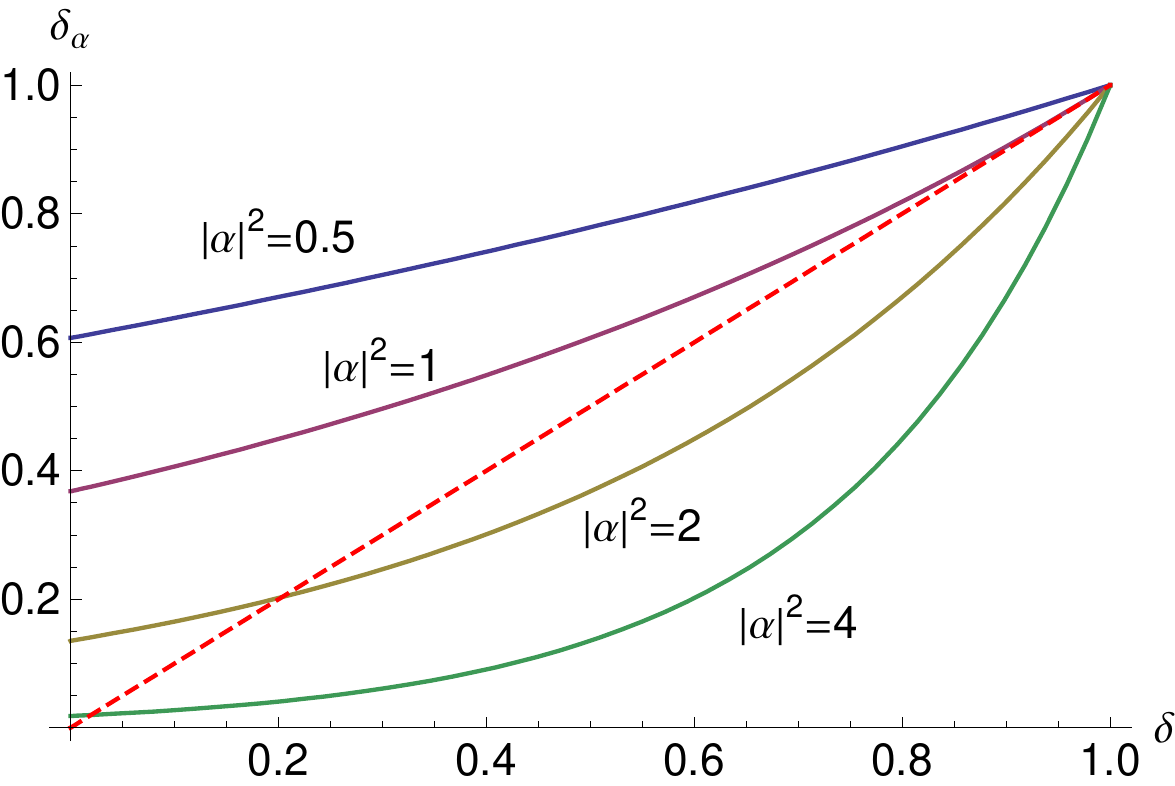}
\caption{(Color online) Overlaps of states in coherent-state protocols for different values of the mean photon number $|\alpha|^2$ and choosing real values of $\delta$ (implying real values of $\delta_{\alpha}$). For $|\alpha|^2<1$, the overlap $\delta_{\alpha}$ is larger than the original overlap $\delta$. For $|\alpha|^2\approx 1$, both the original and coherent-state overlaps are close to each other when $\delta$ is close to 1 and when $|\alpha|^2$ is large, $\delta_{\alpha}$ can be made smaller than almost any value of the original overlap. Finally, for any $\delta\neq 0$ there exists a value of $\alpha$ such that $\delta=\delta_{\alpha}$.}\label{Overlaps}
\end{figure}

Once again, there is an added richness in coherent-state protocols, since the overlaps may be adapted by varying the value of the parameter $\alpha$. For example, in many quantum communication protocols, all overlaps between pairs of states are real numbers, and consequently so are those of their coherent-state versions. In that case, the parameter $\alpha$ can be chosen to increase or decrease the overlap, or to match the exact overlap for a given pair of states. This is illustrated in Fig. \ref{Overlaps}. 

Now that we have outlined the properties of coherent-state protocols, we continue by describing how these techniques can be applied in the construction of protocols in quantum communication complexity and quantum digital signatures.

\section{Quantum Communication Complexity}\label{QCC}
Communication complexity is the study of the amount of communication that is required to perform distributed information-processing tasks. This corresponds to the scenario in which two parties, Alice and Bob, respectively receive inputs $x\in\{0,1\}^n$ and $y\in\{0,1\}^n$ and their goal is to collaboratively compute the value of a Boolean function $f(x,y)$ with as little communication as possible \cite{Yao1979}. Although they can always do this by communicating their entire input, the pertaining question in communication complexity is:  What is the minimum amount of communication that is really needed? Likewise, quantum communication complexity studies the case where the parties are allowed to employ quantum resources such as quantum channels and shared entanglement \citep{BrassardQCC,RevModPhys.82.665}. Remarkably, it has been proven that there exist various problems for which the use of quantum resources offer exponential savings in communication compared to their classical counterparts \cite{DJexp,RazProblem,HM-Bar-Yossef,HM-Gavinsky,QuantumFingerprinting}. In this section, our goal is to employ the mapping to construct protocols that can be implemented using only coherent states and linear optics.

We focus on the bounded-error model in which Alice and Bob have randomness at their disposal and only need to determine the value of the function $f(x,y)$ with probability greater or equal to $1-\epsilon$ (with $\epsilon<\frac{1}{2}$) even for the worst-case values of $x$ and $y$. They can send quantum states to each other, apply unitary transformations on these states, and make measurements in the same way as the quantum communication protocols discussed before. Since they are only interested in learning the value of the function, their final measurement can always be thought of as a projective measurement onto two orthogonal subspaces $H_0$ and $H_1$, corresponding to $f(x,y)=0$ and $f(x,y)=1$ respectively. 

In a coherent-state version of this model, the crucial difference lies in the measurement stage, where the subspaces $H_0$ and $H_1$ are mapped onto sets of modes $S_0$ and $S_1$, where many clicks can occur. In this case, in order to decide between both  values of $f(x,y)$, the strategy is to count the number of clicks that occur in each set of modes. If there are more clicks in the set $S_0$ than in the set $S_1$, the output of the protocol is $f(x,y)=0$, and vice versa. In this way we map the large number of possible click patterns in the coherent-state protocol to the two outcomes of interest.

We now provide conditions such that, if the original protocol had success probability larger than $1-\epsilon$, its coherent-state version will also have success probability larger than $1-\epsilon$. Let $C_b$ be the random variable corresponding to the number of clicks observed in the set of modes $S_b$, with $b=0,1$. The distribution of $C_b$ is known as a Poisson-binomial distribution and its expectation value is given by 
\beq
\mathbb{E}(C_b)=\sum_{k\in S_b}p_{\alpha,k}:=\mu_b.
\eeq 
This distribution can be difficult to work with in its exact form, so it is usual to approximate it by a Poisson distribution with the same mean. This approximation can be made precise through the following result:

\begin{thm}\label{approximation}
 \cite{barbour1992poisson} Let $C_b$ be a Poisson-binomial random variable with mean $\mu_b$. Similarly, let $L_b$ be a Poisson random variable with the same mean $\mu_b$. Then, for any set $A$, it holds that 
\beq
|\Pr(C_b\in A)-\Pr(L_b\in A)|\leq\min(1,\mu_b^{-1})\tau_b,
\eeq
where $\tau_b:=\sum_{k\in S_b}(p_{\alpha,k})^2$ and $p_{\alpha,k}$ is the probability of obtaining a click on
the $k$-th mode.
\end{thm}

We can use this fact to show that, under certain conditions, a coherent-state version of a bounded-error qubit protocol also gives the correct value of the function with bounded error.

\begin{thm}\label{main}
Let a qubit protocol for communication complexity have a probability of success $P_{s}\geq 1-\epsilon$. Then the corresponding coherent-state protocol has a probability of success $P_{\alpha}> 1-\epsilon$ if there exists a mean photon number $\mu=|\alpha|^2$ such that
\begin{align}
2e^{-P_s\mu}(2 e P_s \mu)^{\mu/2}+\max_{\mu_0,\mu_1}\{\min(1,\mu_b^{-1})\}\tau\leq \epsilon\hspace{0.1cm}\label{cond1}
\end{align}
where $\mu_b$ is the expected number of clicks in the set of modes $S_b$ and $\tau=\sum_k(p_{\alpha,k})^2$.
\end{thm}

\textit{Proof.} 
Without loss of generality, we take $f(x,y)=0$ to correspond to the correct value of the function. We can bound the success probability as
\begin{align*}
P_{\alpha}&=\Pr(C_0>C_1)\\
&\geq \Pr(C_0>\tfrac{\mu}{2})\Pr(C_1<\tfrac{\mu}{2})\\
&=(1-\Pr(C_0<\tfrac{\mu}{2}))(1-\Pr(C_1>\tfrac{\mu}{2})).
\end{align*}
From Theorem \ref{approximation} we can also write
\begin{align*}
\Pr(C_0<\tfrac{\mu}{2})\leq \Pr(L_0<\tfrac{\mu}{2})+\min(1,\mu_0^{-1})\tau_0\\
\leq e^{-\mu_0}\left(\frac{2e\mu_0}{\mu}\right)^{\mu/2}+\min(1,\mu_0^{-1})\tau_0,
\end{align*}
where we have bounded the Poisson distribution as in Eq. \eqref{BoundPoisson}. Similarly we have 
\begin{align*}
\Pr(C_1>\tfrac{\mu}{2})\leq e^{-\mu_1}\left(\frac{2e\mu_1}{\mu}\right)^{\mu/2}+\min(1,\mu_1^{-1})\tau_1.
\end{align*}
Putting these together we get
\begin{align*}
P_{\alpha}\geq&\left(1-e^{-\mu_0}\left(\frac{2e\mu_0}{\mu}\right)^{\mu/2}-\min(1,\mu_0^{-1})\tau_0\right)\times\\
&\left(1-e^{-\mu_1}\left(\frac{2e\mu_1}{\mu}\right)^{\mu/2}-\min(1,\mu_1^{-1})\tau_1\right)\\
>& 1-e^{-\mu_0}\left(\frac{2e\mu_0}{\mu}\right)^{\mu/2}-e^{-\mu_1}\left(\frac{2e\mu_1}{\mu}\right)^{\mu/2}\\
&-\min(1,\mu_0^{-1})\tau_0-\min(1,\mu_1^{-1})\tau_1\\
\geq&1-e^{-P_s\mu}(2eP_s\mu)^{\mu/2}-e^{-(1-P_s)\mu}(2e(1-P_s)\mu)^{\mu/2}\\
&-\max_{\mu_0,\mu_1}\{\min(1,\mu_b^{-1})\}\tau,
\end{align*}
where $\tau=\tau_0+\tau_1=\sum_k(p_{\alpha,k})^2$ and we have used the fact that 
\beq
P_s\mu=\sum_{k\in S_0}|\alpha|^2p_k>\sum_{k\in S_0}(1-e^{-|\alpha|^2 p_k})=\mu_0
\eeq 
and similarly $(1-P_s)\mu>\mu_1$. Whenever $P_s>1/2$, it holds that $e^{-P_s\mu}P_s>e^{-(1-P_s)\mu}(1-P_s)$ so we can finally write
\beq
P_{\alpha}> 1-2e^{-P_s\mu}(2eP_s\mu)^{\mu/2}-\max_{\mu_0,\mu_1}\{\min(1,\mu_b^{-1})\}\tau.
\eeq
From this expression it is clear that whenever condition \eqref{cond1} holds, $P_{\alpha}> 1-\epsilon$ as desired. \qed

Notice that the quantity $2e^{-P_s\mu}(2eP_s\mu)^{\mu/2}$ can be made arbitrarily small for any $P_s>1-\epsilon$ by choosing a large enough value of $\mu=|\alpha|^2$. However, large values of $\mu$ result in higher values of the individual click probabilities $\{p_{\alpha,k}\}$, and consequently larger values of $\tau=\sum_k(p_{\alpha,k})^2$, making it harder for the quantity $\max_{\mu_0,\mu_1}\{\min(1,\mu_b^{-1})\}\tau$ to be small. Therefore, condition \eqref{cond1} will only be satisfied when the original probabilities $\{p_i\}$ are very small, as this results in a small $\tau$ even when $\mu$ is large. Of course, whenever the communicated states lie in a Hilbert space of large dimension, we expect the outcome probabilities to be small and the coherent-state protocol to function adequately. 

We are interested in applying the coherent-state mapping to known protocols in quantum communication complexity. In fact, this has already been demonstrated in Ref. \cite{arrazolaqfp}, where, essentially, a coherent-state mapping was used to construct a protocol for quantum fingerprinting. We now discuss how the mapping can be used to construct a protocol for the Hidden Matching Problem.\\

\textit{The Hidden Matching Problem.-} In this communication complexity problem, Alice receives an $n-$bit string $x\in\{0,1\}^n$ as input, with $n$ an even number. Bob receives a matching $M=\{(i_1,j_1),(i_2,j_2),\ldots,(i_{n/2},j_{n/2})\}$ on the set of numbers $\{1,2,\ldots, n\}$, i.e. a partition into $n/2$ pairs. Only one-way communication from Alice to Bob is permitted and their goal is to output at least one element of the matching $(i,j)$ and a corresponding bit value $b$ such that $b=x_i\oplus x_j$, where $x_i$ is the $i$-th bit of the string $x$.

It has been shown that in the bounded-error model, any classical protocol requires $\Omega(\sqrt{n})$ bits of communication \cite{HM-Bar-Yossef}. It was also shown in Ref. \cite{HM-Bar-Yossef} that there exists an efficient quantum protocol that uses only $\order{\log_2 n}$ qubits of communication and outputs a correct answer with certainty. In this protocol, Alice prepares the state
\beq
\ket{x}=\frac{1}{\sqrt{n}}\sum_{i=1}^n(-1)^{x_i}\ket{i}
\eeq
and sends it to Bob, who measures it in the basis 
\beq
\{\tfrac{1}{\sqrt{2}}(\ket{i}\pm\ket{j})\},
\eeq
with $(i,j)\in M$. Since these states form a complete basis, one of these outcomes will always occur, and it will always correspond the correct value since $\tfrac{1}{\sqrt{2}}(\ket{i}+\ket{j})$ only occurs if $x_i\oplus x_j=0$ and similarly, $\tfrac{1}{\sqrt{2}}(\ket{i}-\ket{j})$ only occurs if $x_i\oplus x_j=1$. This allows Bob to give a correct output after performing his measurement. Note that Bob's measurement basis is constructed from the canonical basis by applying a Hadamard transformation to the subspaces $\{\ket{i},\ket{j}\}$, with $(i,j)\in M$.

\begin{figure}
\includegraphics[width=0.9\columnwidth]{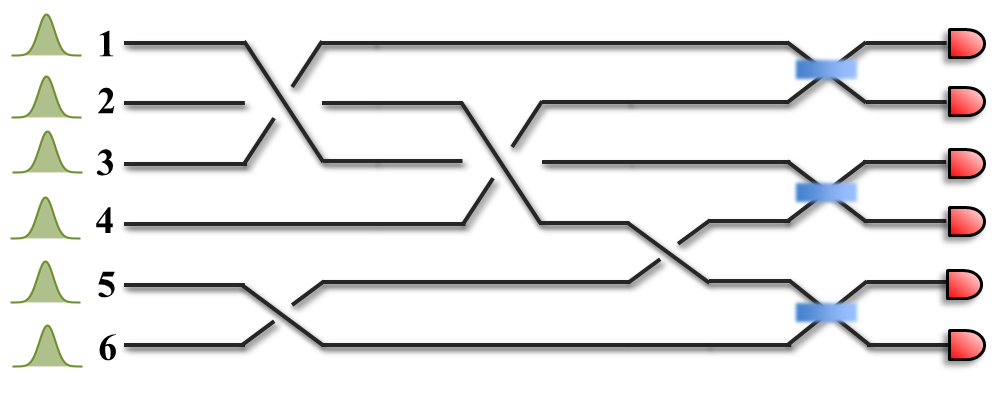}
\caption{(Color online) An example of an implementation of a coherent-state protocol for the Hidden Matching problem. Alice receives a string of six bits and Bob receives the matching $(1,6),(2,5),(3,4)$, as represented in the graph. Alice encodes her input values in the phases of six coherent states in different modes and sends them to Bob. His measurement consists of a circuit in which the modes are permutated in accordance with the matching and then interefere pairwise in three balanced beamsplitters. Bob can output a correct solution to the problem based on the detectors that click.}\label{HM}
\end{figure} 

To construct a coherent-state protocol for the Hidden Matching problem, we just have to apply the rules of the mapping. In this case, Alice prepares the state
\beq
\ket{\alpha, x}=\bigotimes_{i=1}^{n}\ket{(-1)^{x_i}\tfrac{\alpha}{\sqrt{n}}}
\eeq
and sends it to Bob. The linear-optical equivalent of a Hadamard gate is a balanced beam-splitter, so Bob's measurement consists of interfering each of the pairs of modes $\{b_i,b_j\}$ (with $(i,j)\in M$) in a balanced beam-splitter and detecting photons in the outputs as illustrated in Fig. \ref{HM}. If the incoming states to the input ports of the beam splitter are 
\beq
\ket{(-1)^{x_i}\tfrac{\alpha}{\sqrt{n}}}\otimes\ket{(-1)^{x_j}\tfrac{\alpha}{\sqrt{n}}},
\eeq
the output states will be
\beq
\ket{\left(1+(-1)^{x_i\oplus x_j}\right)\tfrac{\alpha}{\sqrt{n}}}\otimes\ket{\left(1-(-1)^{x_i\oplus x_j}\right)\tfrac{\alpha}{\sqrt{n}}}.
\eeq

Notice that for each possible value of $x_i\oplus x_j$, one of the output states will be a vacuum while the other is a coherent-state with non-zero amplitude. Therefore, we can associate a value of $x_i\oplus x_j$ to each of the output detectors so that whenever a click occurs, the correct value can be inferred with certainty. Even if there are many clicks, they will always correspond to a correct value. Thus, the only issue that can arise is that no-clicks occur and the probability that this happens is given by
\beq
P_{\text{no-click}}=e^{-|\alpha|^2},
\eeq
which can be made arbitrarily small by choosing $\alpha$ appropriately. Moreover, Theorem \ref{dimensionthm} guarantees that the amount of information that is transmitted in the coherent-state protocol is $\order{\log_2 n}$ and an exponential separation in communication complexity is maintained. 

Having explored the application of the coherent-state mapping to the realm of quantum communication complexity, we now study how it can be useful in the context of quantum digital signatures.

\section{Quantum Digital Signatures}\label{QDSsection}
Quantum digital signatures (QDS) were first proposed in Ref. \cite{QDS} as a method of guaranteeing the authenticity and integrity of a classic message with unconditional security based only on fundamental principles of quantum mechanics. Recently, other protocols for QDS have been proposed that, notably, use sequences of coherent states and linear optics transformations \cite{dunjko2014quantum,clarke2012experimental,collins2013optical,dunjko2014QDSQKD}. In this section, we apply the mapping to construct a new QDS protocol that can be realized using sequences of coherent states and simple linear optics transformations. In addition to this, we use the mapping to establish a connection between the protocol of Ref. \cite{QDS} with those of Refs. \cite{dunjko2014quantum,clarke2012experimental,collins2013optical} and also to better understand the properties of these latter protocols. 

We begin by briefly describing a simplified version of the QDS protocol of Ref. \cite{QDS}. Alice holds a pair of classical bit strings $\{k_0,k_1\}$ known as private keys, which she will later use to sign a single bit $b$. The protocol makes use of an encoding of classical bit strings into quantum states $k\mapsto\ket{f_k}$ which is known to all recipients. For instance, the encoding could correspond to the states \cite{QuantumFingerprinting}
\beq\label{qfpstates}
\ket{f_k}=\frac{1}{\sqrt{n}}\sum_{i=1}^n(-1)^{k_{b,i}}\ket{i},
\eeq
where $k_{b,i}$ is the $i$-th bit of the private key $k_b$. 

In the distribution stage of the protocol, Alice sends each recipient, Bob and Charlie, two copies of the quantum states corresponding to her private keys and bit value, i.e. she sends $(0,|f_{k_0}\rangle
,|f_{k_0}\rangle)$ and $(1,|f_{k_1}\rangle,|f_{k_1}\rangle)$ to each one. Upon receiving the quantum states from Alice, and for each value of $b$, the recipients perform a swap test on each pair of states to test whether they are different. Once this is done, one of the recipients, for example Bob, forwards one of the copies to Charlie, who also performs a swap test on his state and the forwarded state in order to check that they do not differ. If any of the swap tests fail, the protocol is aborted. 

In the messaging stage, Alice signs a single bit by publicly announcing the message $(b,k_b)$. Each recipient independently performs the map $k_b\mapsto|f_{k_b}\rangle$ and checks, using another swap test, that the resulting state coincides with the one sent by Alice during the distribution stage. Intuitively, security against forging arises from the fact that only an exponentially smaller number of bits can be learned from a measurement of the states $|f_{k_b}\rangle$ compared to the number of encoded bits, and security against repudiation is obtained from the swap tests which prevent Alice from sending differing states to each recipient.

This protocol is difficult to implement, since it requires the preparation and transmission of complex quantum states, performing challenging operations on the states and storing them in a quantum memory until the messaging stage. Instead, we can apply the coherent-state mapping to build a new protocol that is much simpler to implement.

First consider the states of Eq. \eqref{qfpstates} that are used as public keys. Applying the mapping we obtain the states
\beq\label{qdsstates}
|\alpha,f_{k_b}\rangle=\bigotimes_{i=1}^m\ket{(-1)^{k_{b,i}}\tfrac{\alpha}{\sqrt{n}}}_i.
\eeq

When the recipients receive the states, they need to ensure that both copies are equal. Originally, they achieved this by performing a swap test. However, using coherent states there is a simpler alternative: They can pass a single state through a balanced beam splitter, creating copies of the original state, albeit with reduced amplitude. Similarly, even though we could apply the mapping to the operation corresponding to the swap tests between recipients, we can alternatively perform a simpler equality test as was outlined in Ref. \cite{arrazola2013constant} for the case of quantum fingerprinting. This test needs only the interference of the individual coherent states in a balanced beam splitter, and the measurement statistics are analogous to the outcome probabilities of a swap test. Furthermore, as is described in Ref. \cite{dunjko2014quantum}, in a coherent-state protocol it suffices to perform unambiguous state discrimination \cite{PhysRevA.74.022304} of the individual coherent states and store the unambiguous outcomes in a classical memory. This classical data can then be used to verify the authenticity of the signed message. Overall, the QDS protocol is specified as follows:\\
\begin{figure}
\includegraphics[width=\columnwidth]{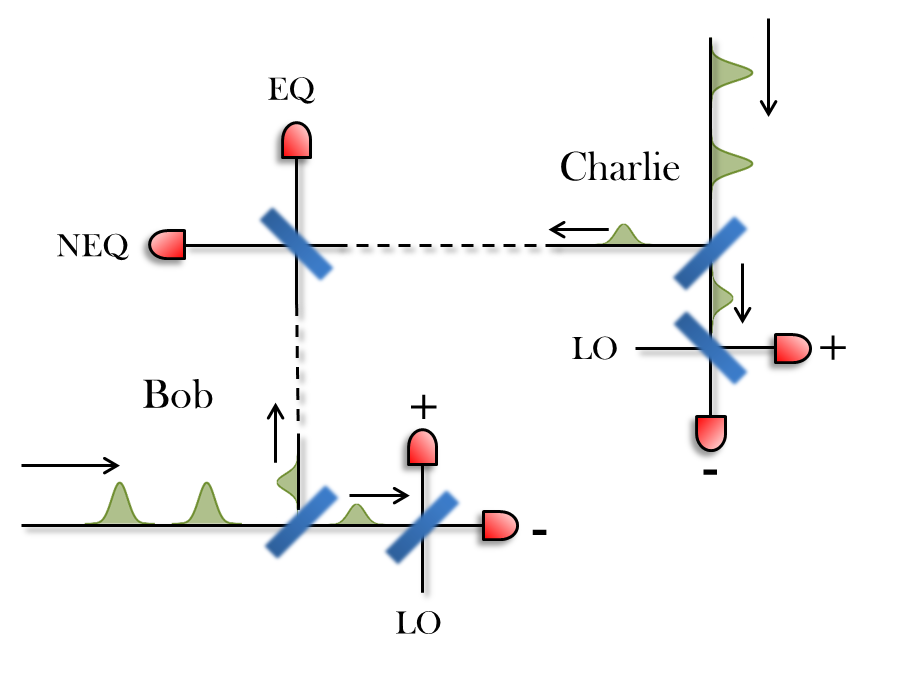}
\caption{(Color online) A protocol for QDS. Bob and Charlie receive a train of pulses from Alice, which they split into two copies of smaller amplitude, keeping one copy for themselves and interfering the other in a balanced beam splitter in Bob's lab. They perform unambiguous state discrimination of the states $|\tfrac{+\alpha}{\sqrt{2n}}\rangle$ and $|\tfrac{-\alpha}{\sqrt{2n}}\rangle$ by interfering their kept signals with a local oscillator (LO). Whenever it is unambiguously revealed that the incoming state could not have been $|\tfrac{+\alpha}{\sqrt{2n}}\rangle$ or $|\tfrac{-\alpha}{\sqrt{2n}}\rangle$, the information is stored in a classical memory. Simultaneously, Bob and Charlie compare their received pulses by interfering their signals and they record the number of ``Not Equal" (NEQ) outcomes. If they exceed a certain fraction of all clicks, the protocol is aborted.}\label{QDSFig}
\end{figure}

\begin{center}
\textbf{QDS Protocol}
\end{center}
\begin{enumerate}
\item{Alice selects two $n$-bit strings $k_0,k_1$ uniformly at random and sends each of Bob and Charlie a copy of the states
\begin{align*}
|\alpha,f_{k_b}\rangle=\bigotimes_{i=1}^m\ket{(-1)^{k_{b,i}}\tfrac{\alpha}{\sqrt{n}}}_i.
\end{align*}
She sends a copy of each to both Bob and Charlie.}
\item{Each recipient passes the states through a balanced beam splitter, obtaining two copies of the states $|\tfrac{\alpha}{\sqrt{2}},f_{k_b}\rangle$.}
\item{Using the first copy of their state, Bob and Charlie perform an unambiguous state discrimination of the individual coherent states $|+\tfrac{\alpha}{\sqrt{2n}}\rangle, |-\tfrac{\alpha}{\sqrt{2n}}\rangle$ and store the unambiguous outcomes in a classical memory.}
\item{One of the recipients sends his second copy to the other and both of these second copies are interfered in a balanced beam splitter, whose output ports are labelled `Equal' and `Not Equal'. If more than a certain fraction $f$ of clicks are observed in the `NEQ' detector, the protocol is aborted.}
\item{To sign a message, Alice publicly reveals $(b,k_b)$ and sends it to one of the recipients, Bob for example. Bob accepts the message as valid if the number of mismatches between his unambiguous outcomes and Alice's revealed key is below a certain fraction $s_a>0$.}
\item{To forward the message to Charlie, Bob sends him Alice's message. Charlie verifies the validity of the message if the number of mismatches between his unambiguous outcomes and Alice's revealed key is below a certain fraction $s_v>s_a$. }
\end{enumerate}

Overall, the QDS protocol we have constructed through the coherent-state mapping can be implemented with the use of sequences of coherent states and beam splitters only, as illustrated in Fig. \ref{QDSFig}. 

It is important to note that we are not providing a full security proof for this protocol, since this is usually a demanding and lengthy task, and the security statements can be complicated functions of the many protocol parameters. Our main goal in this section is to illustrate the usefulness of the mapping to construct and understand quantum communication protocols. Nevertheless, we expect that a full security proof can be constructed from the results of Refs. \cite{QDS} and \cite{dunjko2014quantum}.

The coherent-state mapping can also be used to understand existing QDS protocols. For example, in the protocol of Ref. \cite{dunjko2014quantum}, the public-key states are also of the form of Eq. \eqref{qdsstates}. This implies that they can be equally thought of as arising from an application of the coherent-state mapping to the qubit states of Eq. \eqref{qfpstates}. Furthermore, we note that the protocol of Ref. \cite{dunjko2014quantum} has already been experimentally demonstrated \cite{collins2013optical}, showing that these protocols can indeed be readily implemented.

The mapping also helps us to understand why these QDS protocols do not require a quantum memory. In a qubit protocol for QDS, only $\log_2 n$ bits of information of the private key can be obtained from a measurement on the states of Eq. \eqref{qfpstates}. On the other hand, as discussed in section \ref{CSProtocols}, by choosing the value of $\alpha$ appropriately, we can effectively choose the amount of information about the private keys that can be obtained. This permits the protocol to enter the statistical domain in which enough classical information can be gathered in order to verify the authenticity of a message, but not enough information is available to successfully forge a message.

\section{Conclusions}
We have outlined a general framework for encoding quantum communication protocols involving pure states, unitary transformations, and projective measurements, into another set of protocols that employs coherent states of light in a superposition of modes, linear optics transformations, and measurements with single-photon threshold detectors. This provides a general method for mapping protocols in quantum communication into a form in which they can be implemented with current technology.

The advantages of the coherent-state protocols obtained from the mapping come at the price of a number of optical modes that is equal to the dimension of the original states in the qubit protocol. For practical purposes, this implies that they are suited for protocols that originally do not require a very large number of qubits. But as we have seen, there exists a regime in which the mapping leads to practical protocols whose implementation was previously inaccessible. As such, we expect that our results will pave the way for the experimental demonstration of a wide range of protocols in quantum communication.

From a theoretical perspective, the coherent-state mapping can be thought of as a tool for understanding fundamental aspects about quantum communication and information. For example, the mapping provides us with a connection between two intrinsically quantum properties: entanglement and non-orthogonality. Additionally, the mapping can be also applied in reverse: obtaining qubit protocols from coherent-state protocols. This may serve as a theoretical test bed for proving results regarding qubit protocols, in the same way as many other dualities have been useful in both physics and mathematics.

J.M. Arrazola would like to thank A. Ignjatovic for her help in preparing this manuscript and designing the figures. He is grateful for the support of the Mike and Ophelia Lazaridis Fellowship. We acknowledge support from Industry Canada, the NSERC Strategic Project Grant (SPG) FREQUENCY and the NSCERC Discovery Program.
\bibliography{References}
\bibliographystyle{apsrev}

\end{document}